# The evolution and structure of biomedical knowledge on cytochrome P450

David Fajardo-Ortiz,[1,2] Miguel Lara[2] and Víctor M. Castano[2]

1. Centro de Investigación en Políticas, Población y Salud, Facultad de Medicina, Universidad Nacional Autónoma de México. Cd. Universitaria, Cd. de México, 04510 MEXICO.
2. Coordinación de Vinculación Institucional, Secretaría de Desarrollo Institucional, Universidad Nacional Autónoma de México Cd. Universitaria, Cd. de México, 04510 MEXICO.

**Abstract**

Cytochrome P450 are fundamental proteins to the metabolism of drugs and other relevant processes. through a combination of text mining and network analysis of the P450 literature we mapped the emergence and evolution of the biomedical research communities working on this family of proteins. Our results suggest that the historical research communities that worked on P450 emerged and were organized mainly around methodological achievements like the induction of animal liver microsomal P450 by drugs, the use of chemical inhibitors of P450 enzymes in in-vitro metabolism studies and the development of E. coli expression systems. We found clear evidence that P450 research indeed constitutes a material scientific culture, as we discuss in the text.

**Introduction**

Cytochrome P450 are hemoproteins that absorb light at 450 nm, when reduced in the presence of carbon monoxide.[1] This family of enzymes is found in all the three domains (Eukarya, Archaea, and Bacteria) and one giant virus species.[2] Moreover, P450 enzymes participate in several metabolic processes, such as those of fatty acid, eicosanoid, and vitamin A and D metabolism, as well as metabolism of xenobiotics like pharmaceuticals and carcinogens.[3] Not surprisingly then, P450 proteins are studied by a plethora of disciplines including cancerology, biochemistry, biotechnology, enzymology, physiology, pharmacology, and toxicology.[4-6] P450 research is a consolidated R&D field with over fifty years of history.[4-6] Despite the diversity of disciplines working on P450, it has been suggested the existence of a large and historically-based research community of P450.[4-6] Scientific leaderships, problems, methodologies, and fundamental achievements have been pointed out for this regard.[4-6] However, it is unknown how this putative research community has been organized, and how this community evolves in terms of literature networks, from a literature network perspective. These questions are quite relevant as P450 research is simultaneously a very specialized, extensive, and diverse research topic. To provide a fresh insight on the relevance of analyzing the network of papers on P450 research, it is important to provide a brief background on the studies on the structure of biomedical research.

**Previous research on the structures and evolution of biomedical knowledge**

Previously, it has been described the structure and evolution of the biomedical and clinical knowledge on specific diseases (Anthrax,[7] cancers,[8, 9] ebola[10] and HIV/AIDS[11]) from a literature network perspective. At the core of these studies there is the methodological concept of research fronts, which are sub-networks of papers whose citations are "dense but between which they are sparse" produced by different research communities that work on specific topics with defined methodologies and achievements.[12] In general, the evolution and organization of the research on diseases is leaded by a reductionist process of specializations.[7, 8, 10-12] That is, the older research front is normally related to the study of the disease at the epidemiological (populations) to a systemic (patient) level.[7, 8, 10-12] This first research front provides the clinical description of the disease, i.e., the explanandum to be explained in a reductionist fashion by the subsequent research fronts.[7, 8, 10-12] The following research fronts tend to be specialized in the study of cellular/biomolecular process, mechanism or structures that conform the biomedical explanations of the disease.[7, 8, 10-12]

P450 biomedical research is quite different to the previously studied instances, as this family of proteins is a central element for the explanation of several physiological phenomena of medical importance (explananda), like drug metabolism, chemical toxicity or diseases such as P450 oxidoreductase deficiency.[3-6] On the other hand, P450 proteins, as object of study, are mainly investigated in terms of diversity,[13] mechanisms and structure.[14] We expected that the organization and evolution of P450 research would be leaded by these two objectives: the search for explanations of macroscopic phenomena via P450 enzymes and the study of the mechanisms, diversity and structure of the different P450 enzymes.

**Objective**

Accordingly, through a combination of text mining and network analysis we attempt to map the emergence and evolution of the biomedical research communities working on P450 proteins.

**Methodology**

The methodology was previously successfully used to analyze the organization of the research fronts in HIV/AIDS,[11] ebola[10] and cancer-nanotechnology[15, 16]:

1. A search of papers on the cytochrome P450 was performed in the Web of Science Core

Collection[17] during March, 2017.The search criteria were the following: TITLE: (P450 or P-450) AND TITLE: (drug or treat* or pharm* or therap* or patient or clinical or human). Timespan: All years. Indexes: SCI-EXPANDED, SSCI, A&HCI, CPCI-S, CPCI-SSH, BKCI-S, BKCI-SSH, ESCI. 6,527 papers were found.

2. A network model was built with the papers found in the Web of Science by using the software HistCite.[18] Then, the network model was analyzed and visualized with Cytoscape.[19] The network model consisted of 4,741 nodes and 31,649 edges. Then, the nodes (papers) with a minimal in-degree of 4 (1,847 nodes) were selected. It is important to mention that this sub-network of 1,847 nodes received the 91 percent of the inter-citations (28,899 of 31,649).

3. Cluster analysis based on the Newman modularity[20] was performed on the core subnetwork using Clust&see,[21] a Cystoscape plug-in. This analysis divided the sub-network of citation in several research fronts or clusters of papers. The network clusters are defined by Newman as "groups of vertices within which connections are dense but between which they are sparse.[20]"

4. The sub-network was displayed by using the "yFiles organic" algorithm[21], which is based on the force-directed layout.

5. The content of the identified research fronts -the abstract of their papers- was analyzed with KH Coder[21], a software for quantitative content analysis.

**Results**

**The network model**

A network model of 1,848 high cited papers on P450 was built (Figure 1). The papers of this model received the 91 percent of the inter-citations (28,899 of 31,649) from the research topic. Clust&see identified 11 Research fronts. One of these fronts, the sixth largest, was found to be too general to be considered relevant, according to our text mining results. The research fronts (Figures 1 and 2) were named according to the results of the text mining (Table 1) and title and abstract of the top three papers with the highest indegree of each research front (Table 1).

Figure 2 offers a synthesized view of the organization and evolution of P450 research. This figure shows the main interactions among the 10 research fronts which are displayed according to the average year of publication of their papers. Our results suggest that the most of P450 research is organized around research instruments (research fronts 1-3, 6 and 7) whereas research front 4, 5, 8, 9 and 10 are organized around the study of specific molecular mechanism.

**Research fronts related to research instruments**

The three biggest research fronts (research fronts 1, 2 and 3) emerged in three well differentiated periods of time (Figure 3), i.e., they are related to three different stages in the history of P450 research. Research front 2, the oldest and second biggest, is related to the induction of P450 and its subsequent purification from rat liver microsomes (Table 1). Research front 1, the biggest (Table 1), whose papers were on average published one decade later (Figure 2 and 3), is related to the use of chemical inhibitors of P450 in in-vitro metabolism studies (Table 1). Research front 3, the most recent among the research fronts, is related to the development of E. coli expression systems of P450 for large-scale isolation to perform structure/function analyses (Table 1). Research front 6 is related to the use of bacteriophage libraries to clone the DNA coding sequence of P450 forms. Finally, research front 7 is related to the use of human hepatocytes as a tool for evaluating new molecular entities as inducers of P450 enzymes.

**Research fronts related to the study of molecular mechanisms**

Research fronts related to the study of molecular mechanisms are smaller and more peripheral than the methodological research fronts. The biggest research front of this class is research front 4 (Table 1) which is related to the study of the role of P450 in the activation of procarcinogens (Table 1). Research front 4 is strongly attached to research fronts 1 and 2 (Figure 2) and emerged in an intermediate period between the fall of research front 2 and the emergence of research front 1 (Figure 2 and 3). Research front 5 is related to the study of P450 enzymes that metabolize arachidonic acid, which is converted into more than 100 eicosanoid metabolites. Research front 8, the second oldest among all the research fronts (Figures 2 and 3), is related to the study of the aromatase enzyme, a member of the cytochrome P450 superfamily, that have an important role in steroidogenesis. Research front 9 is related to the study of the pharmacological activation of anti-cancer drugs by P450 enzymes. Finally research front 10 is related to the study of P450 oxidoreductase (POR) mutations that produce diseases like Antley-Bixler syndrome with abnormal genitalia and POR deficiency, a particular form of congenital adrenal hyperplasia.

**Discussion**

**P450 research as a material scientific culture**

The most relevant implication of our results is the material nature of P450 research as a scientific culture (For a deeper discussion on material scientific cultures please see [23]), ruled by clearly identified conditions of instrumentality ("constrains that delimit the allowable form and function of laboratory machines themselves"[23]). The classical examples of material scientific cultures are the traditions of instruments in microphysics which were detailed studied by Peter Galison, for whom those two cultures are organized around the building and use of bubble chambers to form images that would serve as evidence for a new particle; and those employing counting devises arrayed around the particle collision event itself that gather data to make statistical arguments for the existence of a particle.[24, 25] In the case of the biomedical research on P450 enzymes, the conditions of instrumentality are (1) the induction of P450 enzymes in animal model livers, (2) the use of chemical inhibitors in vitrometabolism studies and (3) the use of heterologous expression systems to produce human P450 enzymes (See table 1). These conditions of instrumentality in P450 research set out the conditions under which the instruments will be judged or assessed. For example, the criteria that establish under what conditions vitrometabolism, animal model or hepatocytes would more useful to predict specific aspects of the metabolism of a drug candidate in patients. On the other hand, our results identified research fronts fronts 4, 5, 8-10) that are related to the mechanistic explanation of macro/mesoscopic phenomenon (diseases, metabolism of steroids, drugs and arachidonic acid). However, these fronts represent only 29% of our network model.

**How research instruments would lead to the emergence of P450 research fronts**

A useful frame to understand how research instruments would lead to the emergence of the research fronts in P450 research, is the model of scientific progress proposed by Thomas Kuhn.[26] An interpretation of Kunnian ideas is that science is a complex social system that cyclically undergo a sequence of phases, namely: normal science (a period of standardization of the scientific practice and high productivity), anomalies (the standards are questioned and the productivity reaches its maximum peak), crisis (abandon of the standards and a fall in the productivity)  and paradigm shift in which new standards, approached and/or methodologies are established.[27] Figure 3 shows how most of the identified research fronts clearly exhibit stages of rising, stagnation and fall. In the same vein, the succession of the research fronts would be related to paradigm shifts. Particularly, our results suggest that the emergence of the research fronts in P450 research is related to research instruments that opened the doors to new way of investigating P450 proteins. The narrative of how this research instruments conducted to the emergence of research fronts would be as follows:

The research front 2, the oldest, is related to the discovery of P450 which was spectrally detected (but not yet discovered), for the first time, in rat liver microsomes.[28] However, according to our results the class of instruments that led to the emergence of the first research front was the induction of rat liver microsomal P450 by drugs and other chemical compounds, representing an important contribution for the the purification of microsomal P450 (Table 1). This methodology led to the discovery of several P450 isozymes with different metabolism roles.[29] Simultaneously to the drop of front 2, the research front 1 arose (Figures 2 and 3). The class of instruments that explains the emergence of front 1 was the use of chemical inhibitors of P450 enzymes in in-vitro metabolism studies (Table 1). This methodological approach furthered the identification of P450 enzymes involved in the metabolism of drug candidates and the prediction of drug–drug interactions.[30] Knowledge of P450 induction, and inhibition is currently a fundamental part of drug development process.[31] Few years after the emergence of front 1, front 3 emerged (Figures 2 and 3). The emergence of front 3 is derived from the development of E. coli heterologous expression systems to study P450 (Table 1). These instruments allowed to obtain high quantities of P450 and to perform site-directed mutagenesis studies.[32] Importantly, E. coli expression systems allowed the crystallization of eukaryotic P450, which are intrinsic membrane proteins, by expression of P450 enzymes without the trans-membrane leader sequence.[33] These three research fronts represent the 56% of papers of the network model. The successive emergence of these fronts would represent the paradigm shifts in the history of P450 research. On the other hand, the research fronts 6 and 7, which are also related to research instruments (the use of bacteriophage lambda libraries to clone P50 enzymes and the use of human hepatocytes), are much smaller to be related to fundamental changes in the trajectory of P450 research.

**Milestones in P450 research**

Michael R. Waterman, a veteran of P450 research, and his collaborator David C. Lamb elaborated a timeline of the most relevant milestones in P450 research, which include discoveries on bacterial P450 enzymes.[33] It is important to notice that our analysis was focused only on the biomedical research part of this topic. However, a significant part of the milestone pointed by Waterman and Lamb are methodological achievements, instead of discoveries which strength the idea that P450 research is a material scientific culture (organized around instruments classes). Moreover, some of these milestones pointed out by Waterman and Lamb[33] are related to the emergence of research fronts 2 and 3: "Multiplicity of P450s confirmed" in 1975 and 1976 and mainly "resolution of human P450 enzymes" in 1985 and 1986 are related to front 2 whereas "development of E. coli expression systems to produce high levels of P450 for structure/function analysis" in 1991 and "first eukaryotic P450 structure (CYP2C5)" in 2000 are related to front 3.

Interestingly, Furge and Guengerich, the last another veteran of P450 research, agree with our results in terms of what are the big topics and milestone in P450 research: purification of P450 enzymes, inhibition of P450 and the use of expression systems.[4]

**Conclusion**

For the first time, the network structure and dynamics of P450 biomedical research has been elucidated. We showed how well-defined conditions of instrumentality (class of instruments) are at the core of the emergence and evolution of the leading research communities that work on this family of enzymes. Further investigation is needed if a deeper understanding of how these class of research family instruments have evolved and impacted the most recent research on P450, is to be achieved.

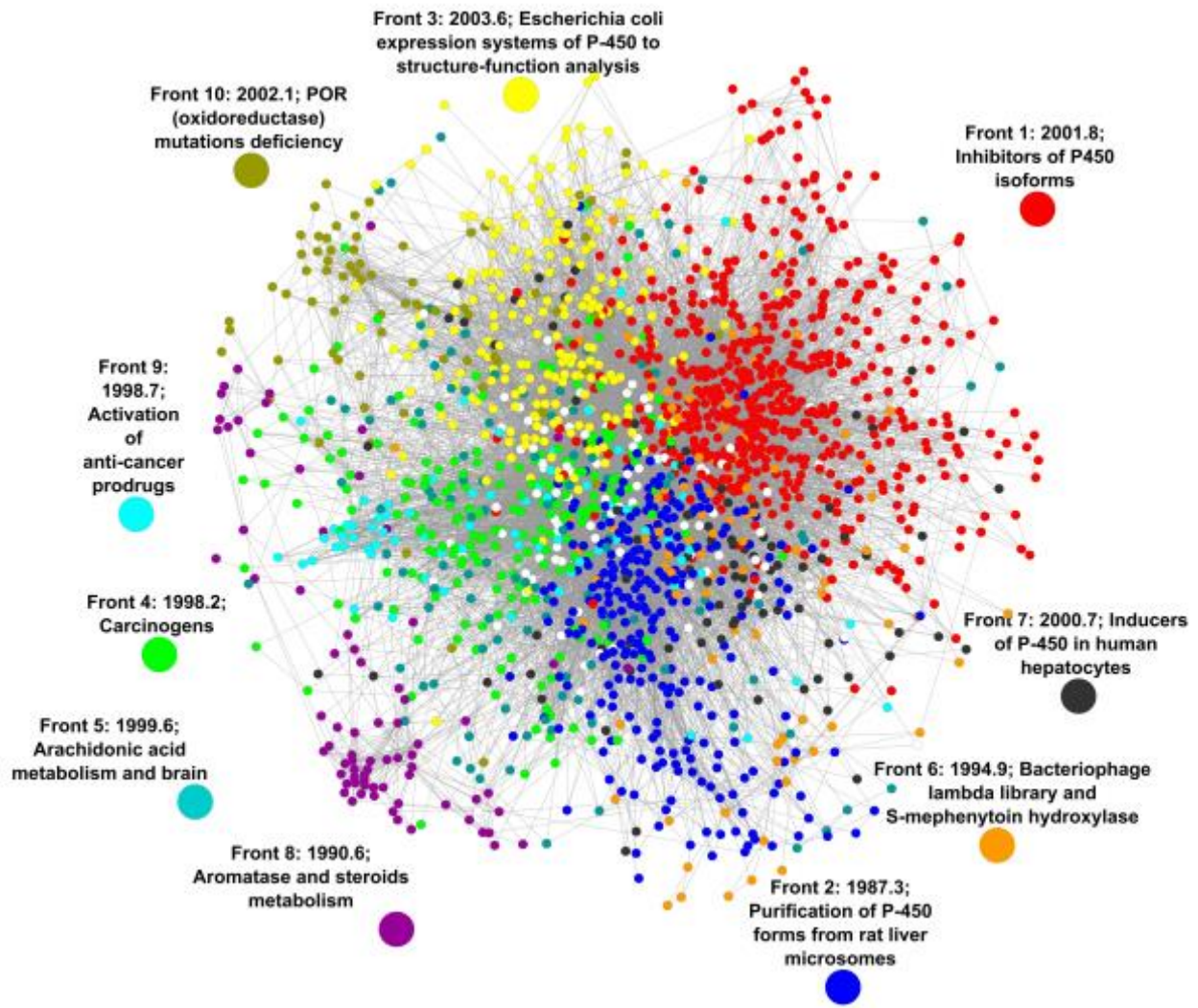

**Figure 1.** The ten research fronts in the network model. The model is displayed by using the "spring embedded" algorithm. The color of nodes (representing the papers) indicates which research front they belong to.

**Figure 2.** Main interactions among the research fronts. Each node represents one of the ten research research fronts. The edges represent the sum of the inter-citations between two research fronts. Only the interactions formed by a minimal of 500 inter-citations or the largest interaction of each research front are shown. Research fronts are displayed according to the average year of publication of their papers.

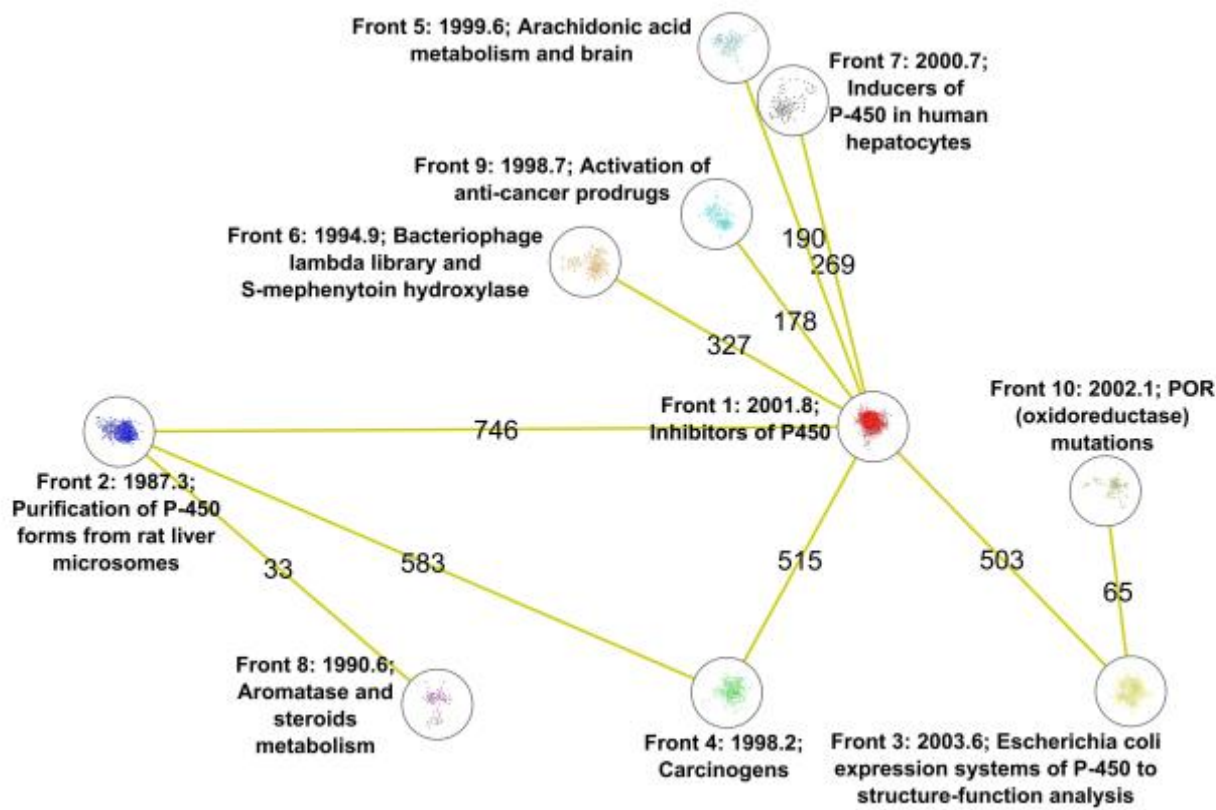

**Figure 3.** Number of papers per year for each of the research fronts.

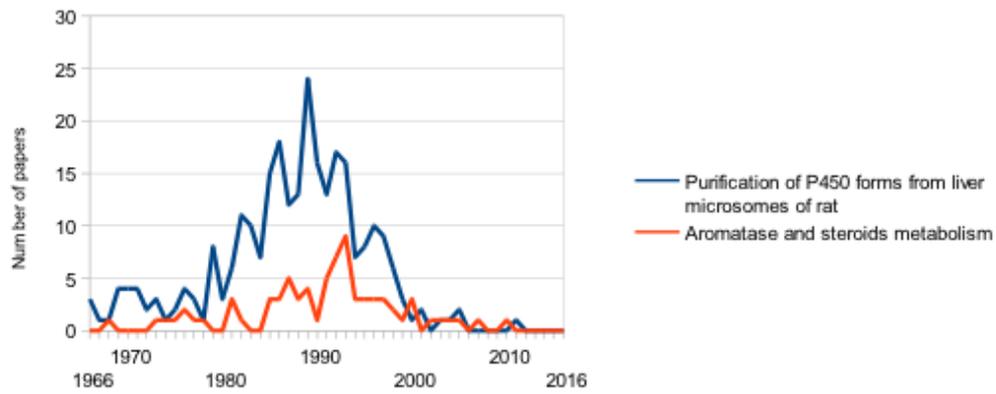

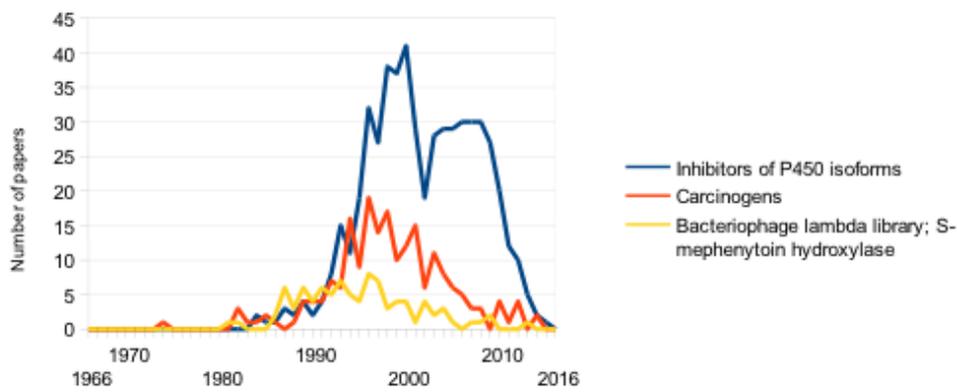

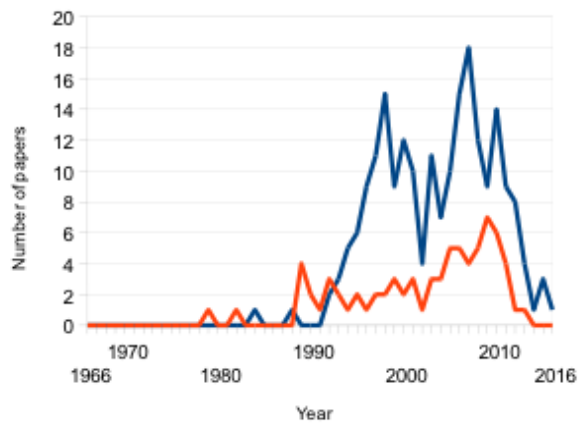

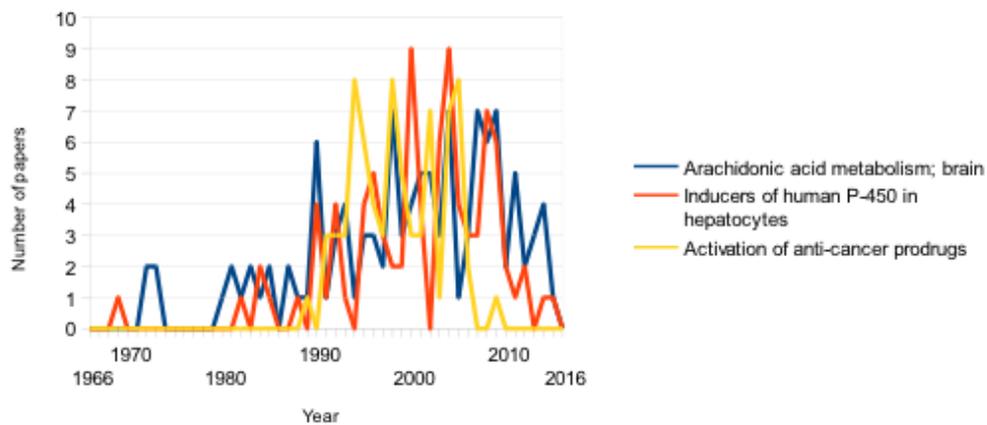

**Table 1.** Top 10 distinctive worlds and list of the three papers with the highest indegree within each research front.

| Front 1 "Inhibitors of P450" ||
|---|---|
| Top 10 most distinctive words | Papers with the highest indegree |
| inhibitor<br><br>CYP<br><br>vitro | Shimada T, Yamazaki H, Mimura M, Inui Y, Guengerich FP. Interindividual variations in human liver cytochrome P-450 enzymes involved in the oxidation of drugs, carcinogens and toxic chemicals: studies with liver microsomes of 30 Japanese and 30 Caucasians. Journal of Pharmacology and Experimental Therapeutics. 1994 Jul 1;270(1):414-23. |
| inhibition<br>microsome<br><br>liver | Newton DJ, Wang RW, Lu AY. Cytochrome P450 inhibitors. Evaluation of specificities in the in vitrometabolism of therapeutic agents by human liver microsomes. Drug Metabolism and Disposition. 1995 Jan 1;23(1):154-8. |
| microM<br>drug<br>CYP2D6<br><br>metabolism | Tassaneeyakul W, Birkett DJ, Veronese ME, McManus ME, Tukey RH, Quattrochi LC, Gelboin HV, Miners JO. Specificity of substrate and inhibitor probes for human cytochromes P450 1A1 and 1A2. Journal of Pharmacology and Experimental Therapeutics. 1993 Apr 1;265(1):401-7. |

| Front 2 "Purification-Induction of P450" ||
|---|---|
| Top 10 most distinctive words | Papers with the highest indegree |
| rat<br>purify<br><br>form | Guengerich FP, Martin MV, Beaune PH, Kremers P, Wolff T, Waxman DJ. Characterization of rat and human liver microsomal cytochrome P-450 forms involved in nifedipine oxidation, a prototype for genetic polymorphism in oxidative drug metabolism. Journal of Biological Chemistry. 1986 Apr 15;261(11):5051-60. |
| gel<br>dodecyl<br><br>purification | Watkins PB, Wrighton SA, Maurel P, Schuetz EG, Mendez-Picon G, Parker GA, Guzelian PS. Identification of an inducible form of cytochrome P-450 in human liver. Proceedings of the National Academy of Sciences. 1985 Sep 1;82(18):6310-4. |
| sodium<br>designate<br>electrophoretic<br><br>p-450nf | Guengerich FP, Dannan GA, Wright ST, Martin MV, Kaminsky LS. Purification and characterization of liver microsomal cytochromes P-450: electrophoretic, spectral, catalytic, and immunochemical properties and inducibility of eight isozymes isolated from rats treated with phenobarbital or. beta.-naphthoflavone. Biochemistry. 1982 Nov 1;21(23):6019-30. |

| Front 3 "E. Coli xpression system" ||
|---|---|
| Top 10 most distinctive words | Papers with the highest indegree |
| active<br><br>coli<br><br>Escherichia<br>bacterial | Gillam EM, Baba T, Kim BR, Ohmori S, Guengerich FP. Expression of modified human cytochrome P450 3A4 in Escherichia coli and purification and reconstitution of the enzyme. Archives of Biochemistry and Biophysics. 1993 Aug 15;305(1):123-31. |
| membrane<br>oxidation<br>site<br>nadph-p-450<br>3A4 | Yano JK, Wester MR, Schoch GA, Griffin KJ, Stout CD, Johnson EF. The structure of human microsomal cytochrome P450 3A4 determined by X-ray crystallography to 2.05-Å resolution. Journal of Biological Chemistry. 2004 Sep 10;279(37):38091-4. |
| | Sandhu P, Guo ZY, Baba T, Martin MV, Tukey RH, Guengerich FP. Expression of modified human cytochrome P450 1A2 in Escherichia coli: stabilization, purification, spectral characterization, and catalytic activities of the enzyme. Archives of Biochemistry and Biophysics. 1994 Feb 1;309(1):168-77. |
| structure | |
| **Front 4 "Carcinogens"** | |
| Top 10 most distinctive words | Papers with the highest indegree |
| carcinogen<br>1A1<br>hydrocarbon<br>pyrene<br>lung<br><br>activation<br>aromatic<br>carcinogenic<br>DNA<br>sample | Guengerich FP, Shimada T. Oxidation of toxic and carcinogenic chemicals by human cytochrome P-450 enzymes. Chemical research in toxicology. 1991 Jul;4(4):391-407.<br><br>Shimada T, Hayes CL, Yamazaki H, Amin S, Hecht SS, Guengerich FP, Sutter TR. Activation of chemically diverse procarcinogens by human cytochrome P-450 1B1. Cancer research. 1996 Jul 1;56(13):2979-84.<br><br>Rendic S, Carlo FJ. Human cytochrome P450 enzymes: a status report summarizing their reactions, substrates, inducers, and inhibitors. Drug metabolism reviews. 1997 Jan 1;29(1-2):413-580. |
| **Front 5 "Arachidonic acid and brain"** | |
| Top 10 most distinctive words | Papers with the highest indegree |
| arachidonic<br>brain<br><br>cortex<br>epoxygenase<br>important<br><br>member<br>cloning<br>endogenous<br>cerebellum<br><br>epoxyeicosatrienoic | Daikh BE, Lasker JM, Raucy JL, Koop DR. Regio-and stereoselective epoxidation of arachidonic acid by human cytochromes P450 2C8 and 2C9. Journal of Pharmacology and Experimental Therapeutics. 1994 Dec 1;271(3):1427-33.<br><br>Ravindranath V, Anandatheerthavarada HK, Shankar SK. Xenobiotic metabolism in human brain—presence of cytochrome P-450 and associated mono-oxygenases. Brain research. 1989 Sep 4;496(1):331-5.<br><br>Nebert DW, Russell DW. Clinical importance of the cytochromes P450. The Lancet. 2002 Oct 12;360(9340):1155-62. |
| **Front 6 "Bacteriophage library"** | |
| Top 10 most distinctive words | Papers with the highest indegree |
| bacteriophage<br>mephenytoin<br>p-450mp<br>tryptic<br>multigene<br><br>hydroxylate<br>tolbutamide<br>closely<br>coding<br>similarity | Umbenhauer DR, Martin MV, Lloyd RS, Guengerich FP. Cloning and sequence determination of a complementary DNA related to human liver microsomal cytochrome P-450 S-mephenytoin 4-hydroxylase. Biochemistry. 1987 Feb 1;26(4):1094-9.<br><br>Beaune PH, Dansette PM, Mansuy D, Kiffel L, Finck M, Amar C, Leroux JP, Homberg JC. Human anti-endoplasmic reticulum autoantibodies appearing in a drug-induced hepatitis are directed against a human liver cytochrome P-450 that hydroxylates the drug. Proceedings of the National Academy of Sciences. 1987 Jan 1;84(2):551-5.<br><br>Ged C, Umbenhauer DR, Bellew TM, Bork RW, Srivastava PK, Shinriki N, Lloyd RS, Guengerich FP. Characterization of cDNAs, mRNAs, and proteins related to human liver microsomal cytochrome P-450 S-mephenytoin 4-hydroxylase. Biochemistry. 1988 Sep 1;27(18):6929-40. |
| **Front 7 "Induction of P450 in hepatocytes"** | |
| Top 10 most distinctive words | Papers with the highest indegree |
| hepatocyte<br><br>inducer<br><br>maintain<br>rifampicin<br>culture<br><br>primary<br>induction<br>day<br>phenobarbital<br><br>prototypical | Pichard L, Fabre IS, Fabre GE, Domergue JA, Saint Aubert BE, Mourad GE, Maurel PA. Cyclosporin A drug interactions. Screening for inducers and inhibitors of cytochrome P-450 (cyclosporin A oxidase) in primary cultures of human hepatocytes and in liver microsomes. Drug Metabolism and Disposition. 1990 Sep 1;18(5):595-606.<br><br>Diaz D, Fabrev I, Daujat M, Saint Aubert B, Bories P, Michel H, Maurel P. Omeprazole is an aryl hydrocarbon-like inducer of human hepatic cytochrome P450. Gastroenterology. 1990 Sep 30;99(3):737-47.<br><br>Madan A, Graham RA, Carroll KM, Mudra DR, Burton LA, Krueger LA, Downey AD, Czerwinski M, Forster J, Ribadeneira MD, Gan LS. Effects of prototypical microsomal enzyme inducers on cytochrome P450 expression in cultured human hepatocytes. Drug Metabolism and Disposition. 2003 Apr 1;31(4):421-31. |
| **Front 8 "Aromatase and steroid metabolism"** | |
| Top 10 most distinctive words | Papers with the highest indegree |
| aromatase<br>placental<br>estrogen<br><br>aromatization<br><br>androstenedione<br><br>initiation<br>hydroxyandrostenedione<br>p-450arom<br>P-450AROM<br>begin | Thompson EA, Siiteri PK. The involvement of human placental microsomal cytochrome P-450 in aromatization. Journal of Biological Chemistry. 1974 Sep 10;249(17):5373-8.<br><br>Mendelson CR, Wright EE, Evans CT, Porter JC, Simpson ER. Preparation and characterization of polyclonal and monoclonal antibodies against human aromatase cytochrome P-450 (P-450AROM), and their use in its purification. Archives of biochemistry and biophysics. 1985 Dec 1;243(2):480-91.<br><br>Corbin CJ, Graham-Lorence S, McPhaul M, Mason JI, Mendelson CR, Simpson ER. Isolation of a full-length cDNA insert encoding human aromatase system cytochrome P-450 and its expression in nonsteroidogenic cells. Proceedings of the National Academy of Sciences. 1988 Dec 1;85(23):8948-52. |
| **Front 9 "Activation of anti-cancer prodrugs"** | |
| Top 10 most distinctive words | Papers with the highest indegree |
| prodrug<br>cyclophosphamide<br><br>cytotoxic<br><br>cancer<br><br>tumor<br><br>therapy<br>CPA<br>bystander<br>Intratumoral<br>substantial | Chang TK, Weber GF, Crespi CL, Waxman DJ. Differential activation of cyclophosphamide and ifosphamide by cytochromes P-450 2B and 3A in human liver microsomes. Cancer research. 1993 Dec 1;53(23):5629-37.<br><br>Chen L, Waxman DJ. Intratumoral activation and enhanced chemotherapeutic effect of oxazaphosphorines following cytochrome P-450 gene transfer: development of a combined chemotherapy/cancer gene therapy strategy. Cancer Research. 1995 Feb 1;55(3):581-9.<br><br>Wei MX, Tamiya T, Chase M, Boviatsis EJ, Chang TK, Kowall NW, Hochberg FH, Waxman DJ, Breakefield XO, Chiocca EA. Experimental tumor therapy in mice using the cyclophosphamide-activating cytochrome P450 2B1 gene. Human gene therapy. 1994 Aug 1;5(8):969-78. |
| **Front 10 "POR mutatoins"** | |
| Top 10 most distinctive words | Papers with the highest indegree |
| POR<br><br>oxidoreductase | Arlt W, Walker EA, Draper N, Ivison HE, Ride JP, Hammer F, Chalder SM, Borucka-Mankiewicz M, Hauffa BP, Malunowicz EM, Stewart PM. Congenital adrenal hyperplasia caused by mutant P450 oxidoreductase and human androgen synthesis: analytical study. The Lancet. 2004 Jun 26;363(9427):2128-35. |
| steroidogenesis<br><br>mutation<br><br>electron<br><br>syndrome<br><br>Antley-Bixler<br><br>mutant<br><br>FAD<br><br>genitalium | Huang N, Pandey AV, Agrawal V, Reardon W, Lapunzina PD, Mowat D, Jabs EW, Van Vliet G, Sack J, Flück CE, Miller WL. Diversity and function of mutations in p450 oxidoreductase in patients with Antley-Bixler syndrome and disordered steroidogenesis. The American Journal of Human Genetics. 2005 May 31;76(5):729-49.<br><br>Fukami M, Horikawa R, Nagai T, Tanaka T, Naiki Y, Sato N, Okuyama T, Nakai H, Soneda S, Tachibana K, Matsuo N. Cytochrome P450 oxidoreductase gene mutations and Antley-Bixler syndrome with abnormal genitalia and/or impaired steroidogenesis: molecular and clinical studies in 10 patients. The Journal of Clinical Endocrinology & Metabolism. 2005 Jan 1;90(1):414-26. |